\documentstyle[12pt,a4]{article}

\begin{document}

\typeout{--- Title page start ---}

\renewcommand{\thefootnote}{\fnsymbol{footnote}}

\typeout{--- preprint number and date ---}
\begin{flushright}
Imperial/TP/93-94/7

hep-ph/9311224
\end{flushright}
\vskip 1cm

\begin{center}

{\bf Why the real time formalism doesn't factorise\footnote{Talk
presented at the 3rd workshop on thermal fields and their
applications, Banff, Canada.}}

\vskip 1.2cm
{\large\bf A.C.Pearson\footnote{E-mail: A.Pearson@IC.AC.UK}}\\
Blackett Laboratory, Imperial College, Prince Consort Road,\\
London SW7 2BZ  U.K. \\
\end{center}

\vskip 1cm
\begin{center}
{\large\bf Abstract}
\end{center}

We show that in the real time formalism, the generating
functional for thermal Green functions does not factorise. However
for most calculations, the normal real time Feynman rules can still
be used to give correct results.

\vskip 1cm

In this talk, I shall be concerned with the Real Time Formalism
(RTF) of equilibrium thermal field theory as described using path
integral techniques\cite{LvW,NS,Ray}. In particular, I would like to
examine whether or not the partition function factorises into two
pieces when using the RTF. This question is crucial to the RTF as it
is precisely this factorisation which allows us to describe thermal
effects in this formalism using thermal field doublets. Without
factorisation we are forced to consider all the real time contour in
closer detail\cite{EP2} or to use another real time
contour\cite{EP2,NRTF}.

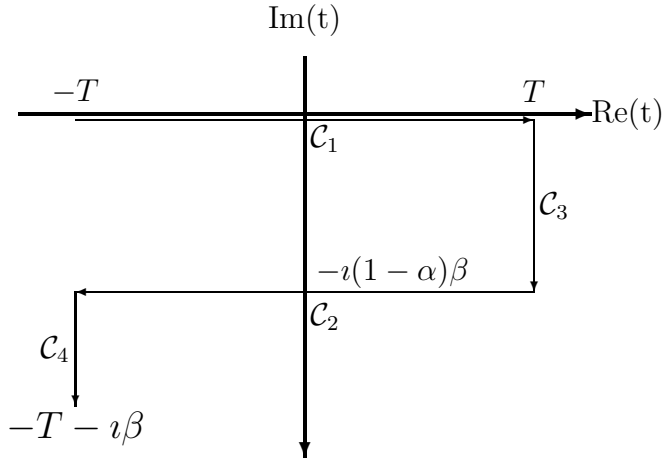
\begin{figure}[htb]
\begin{center}
\setlength{\unitlength}{.3in}
\begin{picture}(0,0)


\thicklines
\put(-5,-1){\vector(1,0){10}}
\put(0,0){\vector(0,-1){7}}


\thinlines
\put(-4,-1.1){\vector(1,0){8}}
\put(4,-1.1){\vector(0,-1){3}}
\put(4,-4.1){\vector(-1,0){8}}
\put(-4,-4.1){\vector(0,-1){2}}
\put(4,-0.6){\makebox(0,0){$T$}}
\put(-4,-0.6){\makebox(0,0){$-T$}}
\put(0.2,-3.9){\makebox(0,0)[bl]{$-\imath(1-\alpha) \beta$}}
\put(-4,-6.2){\makebox(0,0)[t]{\large $-T-\imath \beta$}}
\put(0.1,-1.2){\makebox(0,0)[tl]{${\cal C}_{1}$}}
\put(0.1,-4.3){\makebox(0,0)[tl]{${\cal C}_{2}$}}
\put(4.1,-2.6){\makebox(0,0)[l]{${\cal C}_{3}$}}
\put(-4.1,-5.1){\makebox(0,0)[r]{${\cal C}_{4}$}}
\put(0,0.5){\makebox(0,0)[b]{Im(t)}}
\put(5,-1){\makebox(0,0)[l]{Re(t)}}

\end{picture}
\end{center}
\vspace{2in}
\caption{Path used for the Real Time Formalism.($T\rightarrow\infty$)}
\label{fig:RTF}
\end{figure}

I would like to begin by giving a brief description of what we mean
by factorisation and the key reasons for our desire to split up the
partition function in this way. To do this I shall use a single
scalar field as a simple example. The generating functional of
thermal Green functions is given by
\begin{eqnarray} Z[J] & = & \exp
\bigg{(} -\imath \int_{C} V[-\imath \frac{\delta}{\delta J}]
\bigg{)} . Z_{0}[J] \nonumber \\ Z_{0}[J] &=& \exp \bigg{(}
\frac{-\imath}{2} \int_{C} dt \int_{C} dt' J(t) \Delta_{C}(t-t')
J(t') \bigg{)} \label{partition}
\end{eqnarray}
where $\Delta_{C}(t-t')$ satisfies $(\Box + m^{2}) \Delta_{C}(t,t')
= -\delta_{C}(t-t')$ subject to the KMS condition \cite{KMS}, and $V[\phi]$ is
the interaction potential. I have suppressed spatial indices for
notational convenience. Thermodynamic information may be obtained
from $Z[J=0]$ which is the partition function. The curve, C in the
complex time plane is the path associated with the Real Time
Formalism (see figure \ref{fig:RTF}).

\typeout{figure: Standard RTF curve}

The motivation behind the development of the RTF was to obtain a
convenient means of extracting dynamical information. The usual
imaginary time formalism (ITF) \cite{Ray,ITF} was difficult to use
for this type of calculation as an analytic continuation of the
Euclidean green functions to real times was required. Looking at
figure \ref{fig:RTF} we see that the sections, $C_{1}$ and $C_{2}$
run parallel to the real time axis. As such we can describe the
contributions from these sections in terms of a real time parameter.
However the vertical sections, $C_{3}$ and $C_{4}$, are more
difficult to deal with. If we could ignore the contributions from
$C_{3}$ and $C_{4}$ we could use only real time arguments in this
formalism. The Green functions obtained in this way would also
depend only on real times and so unlike the ITF no analytic
continuation would be required.

In order to be able to `ignore' $C_{3}$ and $C_{4}$ we must be able
to separate their contributions to the generating functional from
the those of $C_{1}$ and $C_{2}$; i.e. we would like to factorise the
generating functional
\begin{equation}
Z[J]=Z_{12}[J].Z_{34}[J] \label{factored}
\end{equation}
where in $Z_{ab}$, all the fields and sources are constrained to lie
on $C_{a}\oplus C_{b}$. Note that we need only consider whether
$Z_{0}$ in Eq.(\ref{partition}) factorises. The interaction term in
$Z[J]$ automatically factorises. Examining $Z_{0}$ in closer detail
we require
\begin{equation}
\int_{C_{a=1,2}} dt \int_{C_{b= 3,4}} dt' J(t) \Delta_{C}(t,t') J(t') = 0
\label{factor}
\end{equation}

To analyse Eq.(\ref{factor}), we shall write the free propagator for
the Klein-Gordon field in its spectral form derived by Mills
\cite{mills}.
\begin{equation}
\Delta_{C}(t-t',\omega)=
\int^{\infty}_{-\infty} \frac{dp_{0}}{2\pi} \rho_{0}(p_{0})e^{-\imath
p_{0} (t-t')} \bigg{[} \Theta_{c}(t-t') + N(p_{0}) \bigg{]}
\label{milleq}
\end{equation}
\begin{eqnarray*}
\rho_{0}(p_{0})=2\pi \bigg{(} \Theta(p_{0}) - \Theta(-p_{0})
\bigg{)} \delta(p^{2}-m^{2}) &, & N(p_{0})=\frac{1}{\exp(\beta
p_{0}) -1}
\end{eqnarray*}

If we could show that the fourier transform of $\Delta_{C}(t-t')$
was a member of the $L^{1}$ class of functions then we could use the
Riemann-Lebesgue lemma on $\Delta_{C}(t-t')$. Looking at
Eq.(\ref{milleq}), we see that the fourier transform of
$\Delta_{C}(t-t')$ contains generalised functions and as such is not
an $L^{1}$ function. However by introducing the
$\epsilon$-prescription we may regulate these
functions\cite{LvW,FS}. By keeping $\epsilon$ finite until the end
of a calculation we may apply the Riemann Lebesgue lemma to show
that $\Delta_{C}(t-t';\epsilon)$ obeys the following rule
\footnote{I have added $\epsilon$ to the arguments of $\Delta_{C}$
to show that its form is altered by the introduction of the
$\epsilon$-prescription.}

\begin{equation}
\lim_{|t-t'| \rightarrow \infty} \Delta_{C}(t-t';\epsilon)=0
\label{infty}
\end{equation}
As an example, we take $t\in C_{1}, t'\in C_{3}$.In this case
$|t-t'| \rightarrow \infty $ unless $t \rightarrow \infty $. If in
addition we restrict the source terms to satisfy the condition
$\lim_{t \rightarrow \infty , t\in C_{1}} J(t) = 0$, we find that
\begin{equation}
\int_{C_{1}} dt \int_{C_{3}} dt' J(t) \Delta_{C}(t-t') J(t') = 0
\end{equation}
A similar reasoning can be applied for other values of $a$
and $b$ in Eq.(\ref{factor}). $Z_{0}[J]$ will factorise if we impose
the so-called `asymptotic condition' \cite{LvW,NS,FS}
\begin{eqnarray}
\lim_{|t| \rightarrow \infty , t\in C_{1}} J(t)=0  &,&
\lim_{Re(t) \rightarrow \pm\infty , t\in C_{2}} J(t)=0
\end{eqnarray}

\section{Should we use the asymptotic condition?}

There are a number of points raised by the use of the asymptotic
condition. Firstly, it seems strange that we have to constrain the
sources in this way when using the RTF yet no constraint is made on
source terms in other formalisms such as the ITF. Also if we are
considering a system in thermal equilibrium, we would expect the
system to be time independent. Since the asymptotic condition is
manifestly time dependent, it seems inconsistent with the assumption
of thermodynamic equilibrium. Finally, it is unusual to constrain
the generating functional through this condition on the source
term. The whole point of the generating functional is that the
sources are not fixed. In particular, we need to consider
infinitesimal variations in the source terms to evaluate the
derivatives in the interaction term of Eq.(\ref{partition}).

There is in fact quite a simple calculation to show that the RTF
will give the wrong answer if we use the asymptotic condition. To
see this, we shall again consider the propagator. We now choose
$t\in C_{3},\/ t'\in C_{4}$. Since $|t-t'| \rightarrow \infty$, we
can use Eq.(\ref{infty}) to show that $\Delta_{C}(t-t')=0$. This
means that $Z[J]$ may be factorised further into three pieces.

\begin{equation}
Z[J]=Z_{12}[J].Z_{3}[J].Z_{4}[J] \label{otw}
\end{equation}
We now make use of the result\footnote{See N.P. Landsmann \& Ch.G. van Weert,
Eq.(2.4.44)},
\begin{equation}
Z_{12}[J=0]=1 \label{j0}
\end{equation}
Eq.(\ref{otw}) now becomes
\begin{eqnarray}
 & Z[J=0] &=Z_{3}[J=0].Z_{4}[J=0] \nonumber \\
or &\ln Z[J=0] &= \ln Z_{3}[J=0] + \ln Z_{4}[J=0] \label{wrong}
\end{eqnarray}

Since $\ln Z$ is the generating functional of connected diagrams,
Eq.(\ref{wrong}) states that to calculate a given connected vacuum
diagram we need only calculate the two cases where all of the vertices
are on $C_{3}$ or all of the vertices are on $C_{4}$. As an example, I
shall use the diagram shown below which can be considered to be a
contribution to $\ln Z[J=0]$.
\begin{eqnarray}
\setlength{\unitlength}{0.5pt}
\begin{picture}(200,50)(0,0)
\thicklines
\put(60,5){\circle{40}}
\put(160,5){\circle{40}}
\put(80,5){\line(1,0){60}}
\put(85,10){$t$}
\put(130,10){$t'$}
\end{picture}
& = &\frac{\lambda^{2}}{8} \bigg{(} \int d^{3}k \Delta_{C}(t=0,\omega)
\bigg{)}^{2} . \sum_{a=3}^{4} I_{aa} \label{wrong2}
\end{eqnarray}

\begin{equation}
I_{ab}= \int_{C_{a}} dt \int_{C_{b}} dt' \; \imath\Delta_{C}(t-t',\omega=m)
\end{equation}
Evaluating Eq.(\ref{wrong2}) we find that
\begin{eqnarray}
I &=& \frac{\lambda^{2}}{8} \bigg{(}\int d^{3}k \Delta_{C}
(t=0,\omega) \bigg{)}^{2} . \nonumber \\
& &\bigg{[} \frac{\imath\beta}{m^{2}} +
\big{(}1-\exp(-\alpha\beta m)\big{)} \big{(} 1-
\exp(-\{1-\alpha\}\beta m)\big{)} \frac{4}{m^{2}} \bigg{]}
\label{I3+4}
\end{eqnarray}
We can compare this result with the same calculation performed
instead using the imaginary time formalism.
\begin{equation}
I_{ITF}= \frac{\lambda^{2}}{8} \bigg{(} \int d^{3}k \Delta_{C}^{2}
(t=0,\omega) \bigg{)}^{2} . \frac{\imath\beta}{m^{2}} \label{eqITF}
\end{equation}
It can be seen that the RTF not only gives the incorrect result but
its answer depends on the unphysical parameter $\alpha$. To remedy
this problem, one must allow the time integrals in Eq.(\ref{wrong2})
to run over the entire contour associated with the RTF. In addition
to the terms from $t,t'\in C_{3}$ and $t,t'\in C_{4}$, we have
contributions from $t,t'\in C_{1}\cup C_{2}$, and from $t\in
C_{1}\cup C_{2}$ and $t'\in C_{3}\cup C_{4}$ (and vice versa). I
shall denote their contributions by $I_{R}$, and by $I_{MIX}$
respectively. We do not include the contributions from $t\in C_{3},
t'\in C_{4}$ since, in this case $\Delta_{c}(t-t')=0$ by virtue of
Eq.(\ref{infty}).
\begin{eqnarray}
I_{R} &=& I_{11}+I_{12}+I_{21}+I_{22}=\frac{4K}{m^{2}}  \label{IR}\\
I_{MIX} &=& I_{13}+I_{31}+I_{14}+I_{41}+I_{23}+I_{32}+I_{24}+I_{42} =
\frac{-8K}{m^{2}}\label{IMIX}
\end{eqnarray}
where $K=\big{(} 1- \exp\{-\alpha\beta m\} \big{)} \big{(} 1- \exp
\{ -(1-\alpha) \beta m\} \big{)}$. If we add these two terms to
Eq.(\ref{I3+4}) then we find that the RTF gives the same answer as the
imaginary time formalism.

There are a number of points arising from the calculation of these
terms. Firstly, it can be seen that $I_{R}$ is non-zero. Since
$I_{R}$ is the full contribution from the $C_{1}\cup C_{2}$ section
of the RTF contour, we would expect this to be zero because of
Eq.(\ref{j0}). The value of $I_{R}$ clearly indicates that
Eq.(\ref{j0}) is wrong. Also, $I_{MIX}$ is non-zero which indicates
that $C_{3}$ and $C_{4}$  cannot be seperated from $C_{1}$ and
$C_{2}$. This suggests that our use of the asymptotic condition is
incorrect. Finally, we note that if $\alpha=0,1$, then $K=0$. This
in turn means that $I_{R}=0$ and $I_{MIX}=0$ as required. In these
special cases. the generating functional does factorise since
either $C_{3}$ or $C_{4}$ becomes the contour associated with the
ITF. From these points we conclude that the RTF does not factorise
unless $\alpha=0,1$ and that should not use the asymptotic
condition.

\section{The RTF without the asymptotic condition}

If the RTF does not factorise, we are faced with the fact that we
must consider all four sections of the real time contour. The only
simplification we can use in our calculations is  Eq.(\ref{infty}).
With this in mind, I would now like to consider a general Feynman
diagram. This diagram will fall into one of two classes: Vacuum
diagrams and every thing else (i.e. diagrams with at least one
external line).

\subsection{Diagrams with external lines}

Diagrams of this type have real times associated with the external
legs. As such, we may use Eq.(\ref{infty}) to show that we need only
consider the contributions from $C_{1}$ and $C_{2}$. In other words
for diagrams with at least one external line, the RTF behaves as if
it does factorise and the usual Feynman rules using a doublet of
fields applies. I shall be using as a specific example, the diagram
shown below. The results derived here apply in general to these
types of diagrams.

\begin{equation}
\setlength{\unitlength}{1pt}
\begin{picture}(100,20)(0,0)
\thicklines
\put(0,5){\line(1,0){30}}
\put(45,5){\circle{30}}
\put(60,5){\line(1,0){30}}
\put(-5,10){$t_{1}$}
\put(25,10){$t$}
\put(65,10){$t'$}
\put(85,10){$t_{2}$}
\end{picture}
=\frac{-\lambda^{2}}{2} \int_{C}dt \int_{C}dt' \; \imath
\Delta_{C}(t_{1}-t) \; \imath \Delta_{C}^{2}(t-t') \; \imath
\Delta_{C}(t'-t_{2})
\end{equation}

Since $t_{1}$ and $t_{2}$ are real and finite, we may use
Eq.(\ref{infty}) to show that either $\Delta_{C}(t_{1}-t)$ or
$\Delta_{C}(t'-t_{2})$ is zero if $t$ or $t'$ are along $C_{3}$ or
$C_{4}$. This means that we need only consider the curves $C_{1}$
and $C_{2}$ and thatwe can use the normal real-time Feynman rules
to evaluate these types of diagrams.

\subsection{Vacuum diagrams}
For these types of diagrams, there is no such direct simplification
to be made. None of the times are fixed for these diagrams and we
must integrate each time over the entire real time contour. The only
way to get around this is to use the method described in Evans
\cite{TSEz}. As a final point it should be noted that diagrams of
this type are associated with static thermodynamic quantities such as
the partition function. As such, the imaginary time formalism is
much better suited to the evaluation of vacuum diagrams.

\section{What about Evans' real time contour?} 

So far I have talked about the real time formalism described using
the contour shown in fig.\ref{fig:RTF}. Recently, another real time
contour has been suggested by Evans\cite{NRTF}. This contour
contains two sections; one along the entire real time axis, the
second comes back but with an infinitesimal slope downwards so that
we arive at the end a distance $-\imath\beta$ below the starting
point of the curve. If we use this contour, we recover the usual
Feynman rules of the conventional RTF\cite{NRTF}.

However, we have just shown that the normal Feynman rules break down
in certain cases. What has gone wrong is that we have ignored the
infinitesimally small gradient of $C_{n2}$. This gradient is of
the order of $\frac{\beta}{T}$ and as such is negligible unless
large times of the order of $T$ are considered. Unfortunately we
ignore large times because of Eq.(\ref{infty}). We must not use the
$\epsilon$-prescription in this formalism.

\section{Conclusions}

We have seen in this talk that we cannot use the asymptotic
condition with the RTF to factorise the generating functional.
However, we have also seen that for most diagrams of interest we
still only need to use the conventional Feynman rules for the RTF
and that these rules only break down in the case of vacuum diagrams.
Finally, we saw that even using a new real time contour did not
really evade the problems associated with the old real time curve
but merely hid them away in a different place.

\section{Acknowledgements}

I would like to thank the organisers of this conference not only for
putting together such a useful and friendly meeting, but also for
providing financial assistance so that I could attend. I would also
like to thank T. Evans, my supervisor.

\renewcommand{\thefootnote}{\arabic{footnote}}
\setcounter{footnote}{0}

\end{document}